\documentclass[10pt,prc,amsmath,amssymb,twocolumn,aps,showpacs,superscriptaddress,groupedaddress]{revtex4-1}
\usepackage{amsmath}
\usepackage{graphicx}
\usepackage[dvips]{color}
\newcommand{\n}[1]{\ensuremath{|\mathbf{#1}|}}
\newcommand{\ve}[1]{\ensuremath{\mathbf{#1}}}

\newcommand{\be}{\begin{equation}}
\newcommand{\ee}{\end{equation}}
\newcommand{\ba}{\begin{eqnarray}}
\newcommand{\ea}{\end{eqnarray}}

\begin{document}

\title{Neutral current quasielastic (anti)neutrino scattering beyond the Fermi gas model at MiniBooNE and BNL kinematics}

\author{M.V.~Ivanov}
\affiliation{Institute\,for\,Nuclear\,Research\,and\,Nuclear\,Energy,\,Bulgarian\,Academy\,of\,Sciences,\,Sofia\,1784,\,Bulgaria}
\affiliation{Grupo\,de\,F\'{i}sica\,Nuclear,\,Departamento\,de\,F\'{i}sica\,At\'omica,\,Molecular\,y\,Nuclear, Facultad\,de\,Ciencias\,\,F\'{i}sicas,\,Universidad\,Complutense\,de\,Madrid,\,Madrid\,E-28040,\,Spain}

\author{A.~N.~Antonov}
\affiliation{Institute\,for\,Nuclear\,Research\,and\,Nuclear\,Energy,\,Bulgarian\,Academy\,of\,Sciences,\,Sofia\,1784,\,Bulgaria}

\author{M.B. Barbaro}
\affiliation{Dipartimento di Fisica, Universit\`a di Torino and  INFN, Sezione di Torino, Via P. Giuria 1, 10125 Torino, Italy}

\author{C. Giusti}
\affiliation{Dipartimento di Fisica,
Universit\`{a} degli Studi di Pavia and INFN, Sezione di Pavia, via Bassi 6 I-27100 Pavia, Italy}

\author{A. Meucci}
\affiliation{Dipartimento di Fisica,
Universit\`{a} degli Studi di Pavia and INFN, Sezione di Pavia, via Bassi 6 I-27100 Pavia, Italy}

\author{J.A. Caballero}
\affiliation{Departamento de F\'{\i}sica At\'{o}mica, Molecular y Nuclear, Universidad de Sevilla, 41080 Sevilla, Spain}

\author{R. Gonz\'alez-Jim\'enez}
\affiliation{Departamento de F\'{\i}sica At\'{o}mica, Molecular y Nuclear, Universidad de Sevilla, 41080 Sevilla, Spain}

\author{E.~Moya de Guerra}
\affiliation{Grupo\,de\,F\'{i}sica\,Nuclear,\,Departamento\,de\,F\'{i}sica\,At\'omica,\,Molecular\,y\,Nuclear, Facultad\,de\,Ciencias\,\,F\'{i}sicas,\,Universidad\,Complutense\,de\,Madrid,\,Madrid\,E-28040,\,Spain}

\author{J.M. Ud\'{\i}as}
\affiliation{Grupo\,de\,F\'{i}sica\,Nuclear,\,Departamento\,de\,F\'{i}sica\,At\'omica,\,Molecular\,y\,Nuclear, Facultad\,de\,Ciencias\,\,F\'{i}sicas,\,Universidad\,Complutense\,de\,Madrid,\,Madrid\,E-28040,\,Spain}

\begin{abstract}

Neutral current quasielastic (anti)neutrino scattering cross sections on a $^{12}$C target are analyzed using a realistic spectral function $S(p,E)$ that gives a scaling function  in accordance with the ($e,e'$)  scattering data. The spectral function accounts for the nucleon-nucleon (NN) correlations by using natural orbitals (NOs) from the Jastrow correlation method and has a realistic energy dependence. The standard value of the axial mass $M_A= 1.032$~GeV is used in all calculations. The role of the final-state interaction (FSI) on the spectral and scaling functions, as well as on the cross sections is accounted for. A comparison of the  calculations with the empirical data of the MiniBooNE and BNL experiments is performed. Our results are analyzed in comparison with those when NN correlations are not included, and also with results from other theoretical approaches, such as the relativistic Fermi gas (RFG), the relativistic mean field (RMF), the relativistic Green's function (RGF), as well as with the SuperScaling Approach (SuSA) based on the analysis of quasielastic electron scattering.

\end{abstract}


\maketitle

\section{Introduction \label{sec:1}}

The analyses of $y$-scaling (see, \emph{e.g.}~\cite{ncsf01, ncsf02, ncsf03, ncsf04, ncsf05, ncsf06, ncsf07, ncsf08, ncsf09, ncsf10}) and superscaling (based on the $\psi'$-scaling variable) (see, \emph{e.g.}~\cite{ncsf10, ncsf11, ncsf12, ncsf13, ncsf14, ncsf15, ncsf16, ncsf17, ncsf18, ncsf19, ncsf20, ncsf21}) phenomena in inclusive electron scattering on nuclei have induced studies of (anti)neutrino-nucleus scattering on the same basis. This allows one to explore fundamental questions of neutrino reactions and neutrino oscillations in relation to the hypothesis of nonzero neutrino masses~\cite{ncsf22}. The theoretical concept of superscaling (a very weak dependence of the reduced cross section on the momentum transfer $q$ at excitation energies below the quasielastic (QE) peak for large enough $q$ and no dependence on the mass number) has been introduced~\cite{ncsf10, ncsf11} within the relativistic Fermi gas (RFG) model. It has been pointed out in~\cite{ncsf13}, however, that the actual dynamical reason of superscaling is more complex than that provided by the RFG model. This imposes the necessity to consider superscaling in the framework of theoretical methods that go beyond the RFG model. An example is~\emph{e.g.}, the Coherent Density Fluctuation Model (CDFM)~\cite{ncsf23,ncsf24} used in~\cite{ncsf16, ncsf17, ncsf18, ncsf19, ncsf25} within this context.

In~\cite{ncsf26} the analyses of superscaling have been extended to include not only QE processes but also those in which $\Delta$-excitation dominates. The QE– and $\Delta$-region scaling functions $f^\text{QE}(\psi')$ and $f^{\Delta}(\psi')$ have been deduced in~\cite{ncsf26} from phenomenological fits to the data for electron-nuclei scattering cross sections by dividing the latter by appropriate elementary $N\to N$ and $N\to\Delta$ functions, respectively. Therefore they include all the effects of the nuclear dynamics, in particular NN correlations and final-state interactions (FSI), which should be reproduced by reliable nuclear models. For instance, in~\cite{ncsf27, ncsf28} a QE scaling function  with asymmetric shape has been obtained in agreement with the experimental scaling function using the relativistic mean field (RMF) model for the final states.

In order to exploit superscaling for neutrino-nucleus studies, in~\cite{ncsf26} (see also~\cite{ncsf27, ncsf29}) the above procedure has been inverted: the scaling functions have been multiplied by the elementary charged-current (CC) (anti)neutrino cross section to obtain the corresponding CC (anti)neutrino cross sections on nuclei for intermediate to high energies in the same region of excitation. The scaling and superscaling ideas have been carried a step further in~\cite{ncsf30} to include neutral current (NC) (anti)neutrino scattering cross sections from $^{12}$C, namely the reactions $^{12}$C($\nu,p\nu)$X, $^{12}$C(${\bar{\nu}},p{\bar{\nu}})$X involving proton knockout and $^{12}$C($\nu,n\nu)$X, $^{12}$C(${\bar{\nu}},n{\bar{\nu}})$X involving neutron knockout in the QE regime. The CDFM scaling function was applied to analyses of neutral current (anti)neutrino scattering on $^{12}$C with energies of $1$~GeV ($u$-channel inclusive processes) in~\cite{ncsf25}. A number of other theoretical studies have been devoted to both neutral-current (\emph{e.g.}~\cite{ncsf31, ncsf32, ncsf33, ncsf34}) and charged-current (\emph{e.g.}~~\cite{ncsf32, ncsf33, ncsf34, ncsf35, ncsf36, ncsf37, ncsf38, ncsf39, ncsf40, ncsf41}) neutrino scattering on nuclei.

Our interest in the present work concerns the analysis of different experimental data recently obtained on neutrino-nucleus processes in several facilities.  At around $1$~GeV, data are available from the MiniBooNE collaboration, both for CC~\cite{AguilarArevalo:2010zc} and NC~\cite{ncsf42} neutrino-$^{12}$C processes, and previous experiment~\cite{ncsf60}. We note also that recent data for antineutrino-nucleus scattering are reported in~\cite{ncsf44, ncsf45}. As known, the analyses of nuclear effects in neutrino scattering are generally regarded as one of the main sources of systematic uncertainties in oscillation experiments, in particular, when it concerns the understanding of charged-current quasielastic (CCQE) interaction with nucleons bound in the nucleus (\emph{e.g.}~\cite{ncsf46}) in the energy range of around $1$~GeV. For energies of a few GeV, the $\Delta$-resonance excitation becomes equally important~\cite{ncsf47} (but keeping in mind that here the nuclear uncertainties are even larger).

Though the main subject of our work is the neutral current QE neutrino scattering by nuclei, in what follows we would like to note various theoretical models that have been used also for the description of CC processes in connection to their applications to analyses of the NC processes.

The analyses of the CCQE MiniBooNE data have raised questions on the capabilities of the various models to account for the different contributions to the neutrino-nucleus scattering cross sections. The RFG model, in which the shell structure and the nucleon correlations are neglected, gives results for the CCQE neutrino scattering that underestimate the data. The accordance with the data is achieved by increasing the world-average axial mass $M_A$ ($M_A= 1.032$~GeV) to $M_A = 1.35$~GeV. An enhancement of the world-average axial mass is required  also in other models based on the impulse approximation (IA) (\emph{e.g.}~\cite{ncsf46, ncsf48, ncsf49, ncsf50, ncsf51, ncsf52, ncsf53, ncsf54, ncsf55}). This is an indication that models based on the IA may lack important contributions to the processes of neutrino-nucleus scattering. In approaches beyond the IA, ingredients such as two-particle two-hole ($2p-2h$) contributions have been included. In the works~\cite{ncsf56, ncsf57} an approach based on the random-phase approximation, improved by considering relativistic corrections~\cite{ncsf51} led to a good agreement with the MiniBooNE data for both CCQE and neutral current quasielastic (NCQE) scattering, including the double differential CCQE cross section. It was pointed out in~\cite{ncsf52} that the multinucleon contribution may effectively be accounted for by increasing the value of the axial mass. As shown in~\cite{ncsf53, ncsf54c, raul1, raul2}, the RMF approach gives a good description of the shape of the double differential cross section from the MiniBooNE experiment but fails to reproduce its normalization. It has been noted in~\cite{ncsf53} that meson exchange currents (MEC) could reduce the discrepancy. The calculations within the relativistic Green's function (RGF) model~\cite{compee, ncsf48,ncsf62} have provided a good description of the total CCQE and of the (double) differential (CCQE) NCQE MiniBooNE cross sections~\cite{ncsf54, ncsf48, ncsf54a, ncsf54c, ncsf62}. The larger RGF cross sections can be attributed to the overall effect of inelastic channels, such as, for instance, rescattering of the knocked-out nucleon, multinucleon processes, non-nucleonic $\Delta$-excitation, that are recovered in the model by the use of a complex relativistic optical potential to describe FSI and that are not included in other models based on the IA.

The SuperScaling Approach (SuSA) previoulsy discussed has been used for analyses of neutrino-nucleus processes in~\cite{ncsf30, ncsf26, ncsf53, ncsf58, Gonzalez-Jimenez:2014eqa, raul1}. The results of SuSA underestimate by $20$--$30$\% the MiniBoonE data. An updated version of the model (SuSAv2), which incorporates different RMF effects in the longitudinal and transverse channels as well as isospin dependence, yields a milder disagreement ($10$-$15$\%)~\cite{Gonzalez-Jimenez:2014eqa}. The account for MEC increases significantly the results for the cross sections reducing the discrepancy with the data in the case of antineutrino but not so much in the neutrino case.

Multinucleon effects on CC neutrino-nucleus scattering have also been investigated by the use of the Giessen Boltzmann-Uehling-Uhlenback event generator in Ref.~\cite{ncsf59}, where the energy spectra of the knockout nucleons are given in detail.

In parallel, it is also interesting to compare with the older NCQE data from  the Brookhaven BNL E734 experiment~\cite{ncsf60}, corresponding to neutrino kinematics similar to MiniBooNE. We should emphasize that the NC cross sections depend also on the strangeness content of the nucleon, particularly, through the axial form factor. Hence a good control of nuclear effects is also required if one wants to extract information on the strange form factors of the nucleon from NC data.

In~\cite{ncsf46} the differential and total cross sections for energies ranging from a few hundreds of MeV to $100$~GeV have been obtained and compared with the data from the BNL E734, MiniBooNE, and NOMAD (see Ref.~\cite{ncsf43}). It has been concluded in~\cite{ncsf46} that the nuclear effects in NCQE and CCQE scattering seem to be very similar, though, according to the authors, the combined analyses  of the CCQE and NCQE data does not seem to support the contribution of multinucleon final states being large enough to explain the normalization of the MiniBooNE cross sections. It should be mentioned that in~\cite{ncsf46} an effective value of $M_A=1.23$~GeV has been used.

The sensitivity to FSI of NCQE (anti)neutrino scattering cross sections has been investigated in~\cite{ncsf62}, where the RGF cross sections calculated with different parametrizations for the phenomenological relativistic optical potential are compared with the MiniBooNE data. The RGF results obtained with the standard value of the axial mass describe well the NCQE (anti)neutrino scattering data. It was pointed out, however, that the application of the RGF model to the semi-inclusive NCQE scattering can also include contributions of channels which are present in the inclusive but not in the semi-inclusive reaction.

More theoretical approaches were presented in Ref.~\cite{ncsf49}, where an axial mass $M_A = 1.28 \pm 0.05$~GeV and a strangeness $\Delta s = 0.11 \pm 0.36$ was extracted, and in Ref.~\cite{ncsf63} within the transverse enhancement (TE) model~\cite{ncsf64, ncsf65}, where a good agreement between the theoretical results and the NCQE data was obtained, that the authors interpret as a very sizable contribution of $2p-2h$ and $3p-3h$ processes.
Finally, in Ref.~\cite{ncsf66} an approach based on a realistic spectral function was applied to calculate the neutron-knockout (through neutral current scattering) cross sections in the kinematical regime of atmospheric-neutrino interactions in a broad energy range extending to $10$~GeV.

The main aim of this paper is to analyze neutral current QE (anti)neutrino scattering cross sections on a $^{12}$C target using a realistic spectral function $S(p,{\cal E})$ that gives a scaling function in accordance with the ($e,e'$) scattering data. In our previous work~\cite{ncsf67} this approach was applied to calculate CCQE (anti)neutrino scattering on $^{12}$C. The spectral function accounts for the NN correlations by using natural orbitals (NO’s) from the Jastrow correlation method and has a realistic energy dependence. In the calculations the standard value of the axial mass $M_A = 1.032$~GeV is used. As in~\cite{ncsf67}, in the present work the role of FSI on the scaling functions and on the cross sections is taken into account. A comparison of the calculations with the empirical data of the MiniBooNE and BNL experiments is performed.

The theoretical scheme of the work is given in Sec.~\ref{sec:2}, where the method to obtain a realistic spectral function as well as the main relationships concerning the NCQE (anti)neutrino-nucleus reaction cross sections are presented. The results of the calculations and the discussion are given in Sec.~\ref{sec:3}, while the summary of the work and the conclusions are included in Sec.~\ref{sec:4}.

\section{THEORETICAL SCHEME \label{sec:2}}

\begin{figure}[b]
\centering
\includegraphics[width=70mm]{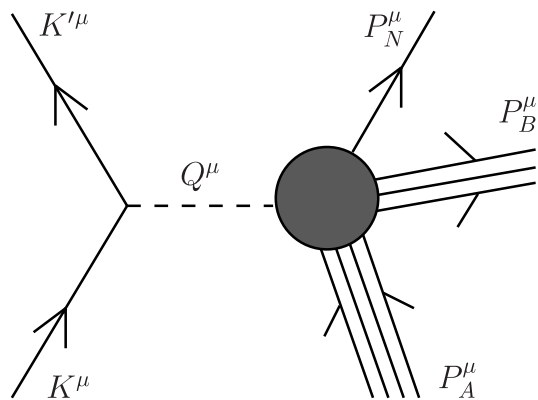}
\caption[]{Kinematics for semi-leptonic nucleon knockout reactions in the one-boson-exchange approximation.\label{kinem}}
\end{figure}

The general formalism for NC (anti)neutrino scattering in the QE regime has been introduced in many previous works~\cite{ncsf30, ncsf33, ncsf79, ncsf70, ncsf71, Alberico:1998qw}. Here we summarize briefly those aspects which are of more relevance for the later discussion of the results and of their comparison with MiniBooNE and BNL data. We consider the semi-leptonic quasi-free scattering from nuclei in Born approximation, assuming that the inclusive cross sections are well represented by the sum of the integrated semi-inclusive proton and neutron emission cross sections~\cite{ncsf30}. The kinematics for semi-leptonic nucleon knockout reactions in the one-boson-exchange approximation is presented in Fig.~\ref{kinem}.

A lepton with 4-momentum $K^{\mu }=(\epsilon ,\mathbf{k})$ scatters to another lepton with 4-momentum $K^{\prime \mu }=(\epsilon ^{\prime },\mathbf{k}^{\prime })$, exchanging a vector boson with $4$-momentum $Q^{\mu }=K^{\mu }-K^{\prime \mu }$. The lepton energies are $\epsilon =\sqrt{m^{2}+k^{2}}$ and $\epsilon ^{\prime } =\sqrt{m^{\prime 2}+k^{\prime 2}}$, where the masses of the initial and final lepton $m$ and $m^{\prime }$ are assumed to be equal to zero for NC neutrino scattering. In the laboratory system the initial nucleus being in its ground state has a $4$-momentum $P_{A}^{\mu } =(M_{A}^{0},0)$, while the final hadronic state corresponds to a proton or neutron with $4$-momentum $P_{N=p~\text{or}~n}^{\mu }=(E_{N},{{\bf p}}_N)$ and an unobserved residual nucleus with $4$-momentum $P_{B}^{\mu }=(E_{B},\mathbf{p}_{B})$. Usually the missing momentum $\mathbf{p} \equiv -\mathbf{p}_{B}$ and the excitation energy $\mathcal{E}\equiv E_{B} - E_{B}^{0}$, with $E_{B}^{0} =\sqrt{\left( M_{B}^{0}\right) ^{2}+p^{2}}$ are introduced, $M_{B}^{0}$ being the ground-state mass of the daughter nucleus. Although in real situations, as is the case for MiniBooNE and BNL, there are usually no monochromatic beams and an integral over the allowed energies folded with the neutrino flux must be performed, we assume the energy of the incident neutrino to be specified and also the outgoing nucleon energy $E_N$ to be known. Finally, the angle $\theta_{kp_N}$ between the incident neutrino and the ejected nucleon momentum is also given.

Starting from the Feynman amplitude associated with the exclusive diagram of Fig.~\ref{kinem}, one can get inclusive cross sections by integrating over the undetected outgoing particles.

In the case of QE electron- or CCQE neutrino-scattering, the outgoing lepton is detected and a sum over the outgoing nucleon variables is performed. Using the language introduced in this context in Ref.~\cite{ncsf79}, we refer to these processes as ``$t$-channel'' reactions, since the Mandelstam variable $t=(K^\mu-K^{\prime\mu})^2$ is fixed. Then the $(e,e^\prime)$ and CC $(\nu_l,l^\prime)$ reactions are governed by the same kinematics and the scaling formalism developed for the former can be trivially extended to the latter.

In the case of NC neutrino scattering only the outgoing nucleon can be experimentally detected, while the unobserved outgoing neutrino is integrated over. This is referred to as a ``$u$-channel'' process, where the Mandelstam variable $u=(K^\mu-P^\mu)^2$ is fixed. Then the kinematics is not the same as in the $(e,e^\prime)$ case and, in particular, the two inclusive cross sections involve an integration over a slightly different region in the missing energy and momentum plane. As a consequence it is not evident that the scaling arguments can be applied to NC scattering. However, in ~\cite{ncsf30} the influence of a non-constant $Q^\mu$ in the derivation of scaling in the NC case was thoroughly investigated within the general framework of the RFG model concluding that the scaling ideas still work properly for NC neutrino-nucleus processes. That study was extended  in~\cite{ncsf71, ncsf72} making use of the RMF approach. These results showed that scaling of the second kind, {\it i.e.,} independence of the nuclear target, works extremely well. On the contrary, scaling of first kind (independence on the transfer momentum) depends on the specific kinematical situation considered. In general, first-kind scaling seems to be well respected when the angle of the ejected nucleon is larger than roughly $50^\circ$. This is the region where the cross section integrated over angles reaches larger values. Therefore, first-kind scaling is expected to work properly at MiniBooNE and BNL. Indeed, for the RMF and the particular kinematics involved in the experiments analyzed in this work (MiniBooNE and BNL), we have verified that the calculation of the NC cross sections based on $u$-scaling (as indicated in Eqs.~(\ref{cr.s.}) and~(\ref{scaling}) plus~(\ref{HF+lorent}) in the text below) gives rise to results very similar (within few percents) to those provided by the full calculation, {\it i.e.,} without resorting to the scaling assumption. This is strictly true for $Q^2 \lesssim 0.7$~GeV$^2$. For larger values of $Q^2$, the $u$-scaling approximation begins to deviate from the full result, but by an amount significantly smaller than the uncertainty linked to the data error bands.

The usual procedure for calculating the $(l,l' N)$ cross section includes the Plane Wave Impulse Approximation (PWIA) and integrations over all unconstrained kinematic variables. It is shown in~\cite{ncsf30} that the inclusive cross section in the $u$-channel can be written after some approximations in the following form:
\begin{equation}\label{Eq01}
\frac{d\sigma}{d\Omega_{N} dp_N} \simeq \overline{\sigma}_{{s.n.}}^{(u)}
F(y',q') ,
\end{equation}
where
\begin{equation}\label{Eq02}
F(y',q') \equiv \int\limits_{{\cal D}_u} p dp \int \frac{d{\cal
E}}{E} \Sigma \simeq F(y'),
\end{equation}
provided the effective NC single nucleon (s.n.) cross section
\begin{multline}\label{Eq03}
\overline{\sigma}_\text{s.n.}^{(u)}= \frac{1}{32\pi
\epsilon}\frac{1}{q^\prime} \left(\frac{p_N^2}{E_N}\right) g^4
\int\limits_0^{2\pi} \frac{d\phi'}{2\pi}l_{\mu\nu}(\mathbf{k},\mathbf{k}')\\
\times w^{\mu\nu}({\mathbf{p}},{\mathbf{p}}_N) D_V(Q^2)^2
\end{multline}
is almost independent of $(p,{\cal E})$ for constant $(k,p_N,\theta_{kp_N})$. In Eq.~(\ref{Eq03}) $l_{\mu\nu}$ and $w^{\mu\nu}$ are the leptonic and s.n. hadronic tensor, respectively, and $D_V(Q^2) $ is the vector boson
propagator~\cite{ncsf30}. In Eq.~(\ref{Eq01}) $Q^{\prime \mu } \equiv K^{\mu }-P_{N}^{\mu } =(\omega ^{\prime },\mathbf{q}^{\prime })$ is the four-momentum transferred from the initial lepton to the ejected nucleon and $y'$ is the scaling variable naturally arising in the $u$-scattering kinematics, analogous to the usual $y$-scaling variable for $t$-scattering. The scaling function $F(y')$ obtained within a given approach can be used to predict realistic NC cross sections. Assuming that the domains of integration ${\cal D}_u$ (in the $u$-channel) and ${\cal D}_t$ (in the $t$-channel) are the same or very similar, the results for the scaling function obtained in the case of inclusive electron scattering (where ${\cal D}_t$ works) can be used in the case of NC neutrino reactions. It is pointed out in~\cite{ncsf30} that ${\cal D}_t$ and ${\cal D}_u$ differ significantly only at large ${\cal E}$ (also at large $p$, but there the semi-inclusive cross sections are expected to be negligible). So, given that the semi-inclusive cross sections are dominated by their behavior at low ${\cal E}$ and low $p$, one expects the results of the integrations in the $t$- and $u$-channel to be very similar, and thus the scaling functions will be essentially the same in both cases.

As noted in~\cite{ncsf30}, if the s.n.~cross section is smoothly varying within the $(p,{\cal E})$ integration region, the differential cross section in the RFG can be factorized as shown in Eq.~(\ref{Eq01}) with the RFG scaling function:
\begin{equation}\label{Eq04}
F_\text{RFG}(\psi_\text{RFG}) = \frac{3}{4} k_F\left(1-\psi^{2}_\text{RFG}\right) \theta\left(1-\psi^{2}_\text{RFG}\right),
\end{equation}
where the RFG $u$-channel $\psi$-variable is defined:
\begin{equation}\label{Eq05}
\psi_{RFG}=s\sqrt{\frac{m_N}{T_F}} \left[\sqrt{1+\left(\frac{y_\text{RFG}}{m_N}\right)^2}-1\right]^{1/2}
\end{equation}
and
\begin{equation}\label{Eq06}
y_{RFG}= s \frac{m_N}{\tau'} \left[ \lambda'\sqrt{\tau'^2\rho'^2+\tau'}-\kappa'\tau'\rho'\right]
\end{equation}
is the RFG $y$-scaling variable for the $u$-channel and corresponds to the minimum momentum required for a nucleon to participate in the NC neutrino-nucleus scattering. The dimensionless kinematic quantities in Eq.~(\ref{Eq06}) are given by: $\kappa'\equiv q'/2 m_N$, $\lambda'\equiv \omega'/2 m_N$, $\tau'=\kappa'^2-\lambda'^2$, $\rho'\equiv 1-\dfrac{1}{4\tau'} (1-m'^2/m_N^2)$. The sign $s$ is
\begin{equation}\label{Eq07}
s\equiv{\rm sgn}\left\{ \frac{1}{\tau'} \left[\lambda'\sqrt{\tau'^2\rho'^2+\tau'}-\kappa'\tau'\rho'\right]\right\}.
\end{equation}

The basic relationships used to calculate the s.n.~cross sections are given in~\cite{ncsf30}. This concerns the leptonic and hadronic tensors and the response and structure functions. Here we summarize the basic expressions for the neutral weak nucleon form factors and the specific set of parameters considered in the calculations:
\begin{itemize}
\item[$\bullet$] Weak mixing angle: $\sin^2\theta_W = 0.23122(15)$,~\cite{ncsf73};
\item[$\bullet$] Isovector axial form factor of the nucleon at zero momentum transfer:
$g_A = 1.2695$,~\cite{ncsf73};
\item[$\bullet$] The weak form factors are given in the form [Eqs.~(55) and (56),
Ref.~\cite{ncsf74}]:
\begin{align}
 &\widetilde{G}_{a}(Q^2)=\xi_V^{T=1}G_{a}^{T=1}\tau_3+
 \sqrt{3}\,\xi_V^{T=0}G_{a}^{T=0}+\xi_V^{(0)}G_a^{(s)}\, ,\notag\\
   &\qquad \qquad \quad a=\{E,M\}, \notag \\
& \widetilde{G}_A(Q^2)=\xi_A^{T=1}G_A^{(3)}\tau_3 + \xi^{T=0}_A G_A^{(8)} +
\xi_A^{(0)}G_A^{(s)},\notag
\end{align}
with (no radiative corrections included)
\begin{gather}
 \xi^{T=1}_V =2(1- 2\sin^2\theta_W)\,; \,\sqrt{3}\,\xi^{T=0}_V = -4\sin^2\theta_W\,;\notag \\
 \xi_V^{(0)} = -1\,; \,\xi^{T=1}_A = -2\,;\, \xi^{T=0}_A = 0\,;\,
 \xi_A^{(0)} = 1\,.\notag
\end{gather}
\item[$\bullet$] The isoscalar and isovector form factors are (for more details,
see~\cite{ncsf74}):
\begin{gather}
G_{E,M}^{T=1} = G_{E,M}^p - G_{E,M}^n,~G_{E,M}^{T=0} = G_{E,M}^p + G_{E,M}^n,\notag\\
G_{A}^{(3)} =  g_A/2\ G_D^A(Q^2),~G_A^{(s)} = g_A^{(s)}\ G_D^A(Q^2),\notag
%G_A^{(8)} &=& (3F-D)/(2\sqrt{3})\ G_D^A(Q^2)\\
\end{gather}
with $G_D^A(Q^2)=(1 + |Q^2|/M_A^2)^{-2}$.
\item[$\bullet$] Electric and magnetic strange form factors (dipole $Q^2$-dependence has
been used):
\begin{eqnarray}
   G_E^{(s)}(Q^2) = \rho_s\tau G_D^V(Q^2),~G_M^{(s)}(Q^2) = \mu_s G_D^V(Q^2).\notag
\end{eqnarray}
\item[$\bullet$] The electromagnetic (EM) form factors correspond to the so-called GKex
(Gari-Kr{\"u}mpelmann extended, developed by E.~L.~Lomon) model~\cite{ncsf75, ncsf76, ncsf77}.

\item[$\bullet$] Unless otherwise specified, the following set of strange parameters has been considered
(see Ref.~\cite{ncsf74} for more details):
\begin{gather}
 \mu_s=-0.020%\pm0.065
 ; \,\rho_s=0.59%\pm0.21
 ; \,g_A^{(s)}=0.\notag
 \end{gather}
 \item[$\bullet$] Finally, the axial mass is taken to be the world-average value $M_A=1.032$~GeV.
\end{itemize}

\begin{figure}[t!]
\includegraphics[width=0.95\columnwidth]{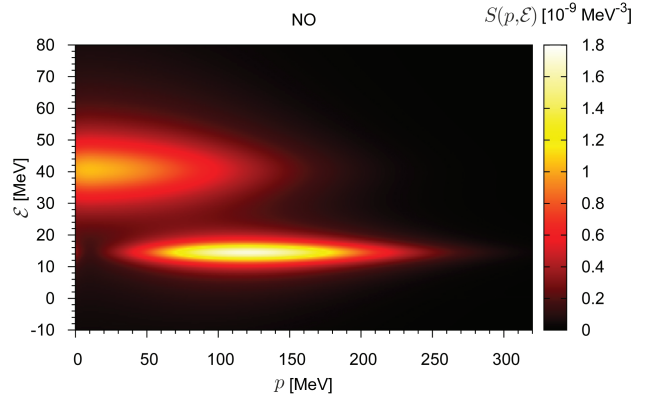}
\caption{(Color online) The $^{12}$C realistic spectral function $S(p,{\cal E})$, which is constructed using natural orbits single-particle momentum distributions from the Jastrow correlation method and Lorentzian function for the energy dependence (see text).}\label{fig00}
\end{figure}

In this work we use in the calculations the RFG, SuSA, harmonic oscillator (HO+FSI) and natural orbitals (NO+FSI) scaling functions including final state interactions (see~\cite{ncsf67} and the text below). We present also results obtained in the RMF and RGF models, where the cross sections are calculated in a fully unfactorized approach which does not make use of the approximations leading to Eq.~(\ref{Eq01}). All our results are used to analyze NCQE (anti)neutrino cross sections on a CH$_2$ target measured by the MiniBooNE collaboration~\cite{ncsf45, ncsf42}. We shall also compare with the BNL E734 experiment~\cite{ncsf60}, studying $\nu p$ and $\overline{\nu} p$  NCQE interactions, where the target was composed in 79\% of protons bound in carbon and aluminum and in 21\% of free protons.

In details, here we present briefly how the HO+FSI and NO+FSI scaling functions are obtained (see also Ref.~\cite{ncsf67}):
\begin{itemize}
\item[(i)] The spectral function $S(p,{\cal E})$ is constructed in the form
\begin{equation}\label{HF+lorent}
    S(p,{\cal E})=\sum_{i}2(2j_i+1)n_i(p) L_{\Gamma_i}({\cal E} - {\cal E}_i);
\end{equation}
\item [(ii)] The single-particle momentum distributions $n_i(p)$ are taken to be either corresponding to the HO single-particle wave functions or to the NO's from the Jastrow correlation method, where central short-range NN correlations are included;
\item [(iii)] The Lorentzian function
\begin{equation}\label{lorent}
L_{\Gamma_i}({\cal E}-{\cal E}_i)=
\dfrac{1}{\pi}\dfrac{\Gamma_i/2}{({\cal E}-{\cal E}_i)^2+(\Gamma_i/2)^2}\, ,
\end{equation}
is used for the energy dependence of the spectral function with parameters $\Gamma_{1p} = 6$~MeV, $\Gamma_{1s} = 20$~MeV, which are fixed to the experimental widths of the $1p$ and $1s$ states in $^{12}$C.

The realistic spectral function $S(p,{\cal E})$ is presented in Fig.~\ref{fig00}, where the two shells $1p$ and $1s$ are clearly visible.

\item[(iv)] For a given momentum transfer $q$ and energy of the initial electron $\varepsilon$ we calculate the electron-nucleus ($^{12}$C) cross section using
    \begin{multline}\label{cr.s.}
        \frac{d\sigma_t}{d\omega d\n q}={2\pi\alpha^2}\frac{\n q}{E_{\ve k}^2}
        \int dE\:d^3p\:\frac{S_t(\ve p, E)}{E_{\ve p}E_{\ve {p'}}}\times\\
       \times        \delta\big(\omega+M-E-E_\ve{p'}\big)L_{\mu\nu}^\text{em}H^{\mu\nu}_{\text{em, }t}\,
    \end{multline}
    in which the spectral function $S(p,{\cal E})$ [Eq.~(\ref{HF+lorent})] is used, the index $t$ denotes the nucleon isospin, $L_{\mu\nu}^{\text{em}}$ and $H^{\mu\nu}_{\text{em, }t}$ are the leptonic and hadronic tensor, respectively (for details, see Ref.~\cite{ncsf67});
\item[(v)] The corresponding scaling function $F(q,\omega)$ is calculated within the PWIA by
 \be F(q,\omega)\cong
\dfrac{\left[d\sigma/d\epsilon'd\Omega'\right]_{(e,e')}}{\overline{\sigma}^{eN}
(q,\omega;p=|y|,{\cal E}=0)} \label{scaling}
\ee
and by multiplying it by $k_F$ the scaling function $f(\psi)$ is obtained, where $k_F$ is the Fermi momentum for a specific nucleus and $\overline{\sigma}^{eN}$ is the azimuthal angle-averaged single-nucleon cross section;
\item[(vi)] To account for FSI, the $\delta$-function in Eq.~(\ref{cr.s.}) is replaced by
    \begin{multline}\label{deltaf}
 \delta (\omega+M-E-E_\ve{p'}) \rightarrow \\ \rightarrow\dfrac{W/\pi}{W^2+[\omega+M-E-E_\ve{p'}-V]^2},
    \end{multline}
where the real ($V$) and imaginary ($W$) parts of the OP are obtained in Ref.~\cite{ncsf78} from the Dirac OP.
\end{itemize}

\section{Results and discussion \label{sec:3}}

\begin{figure}[t]
\includegraphics[width=0.95\columnwidth]{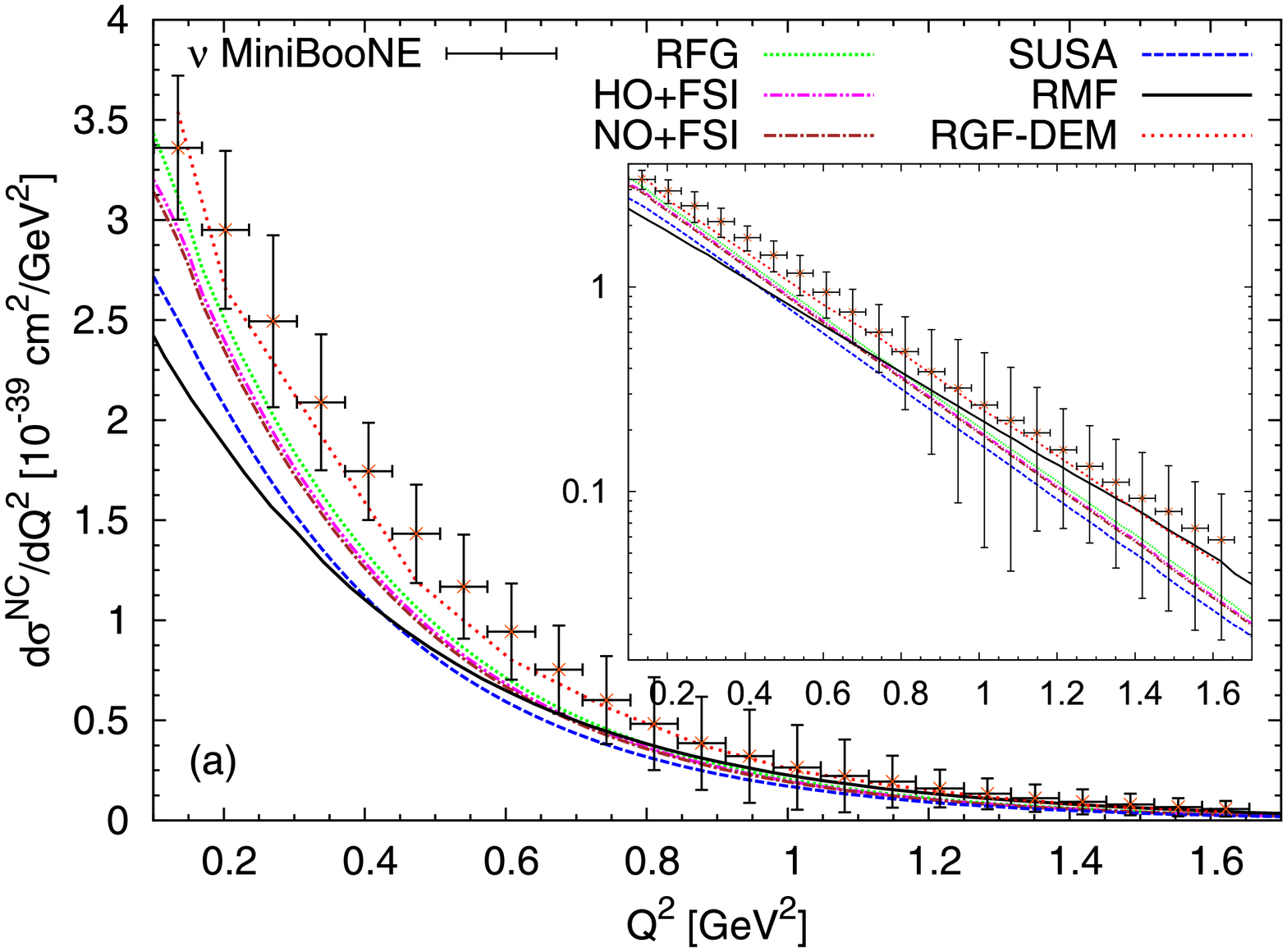}\\[5pt]
\includegraphics[width=0.95\columnwidth]{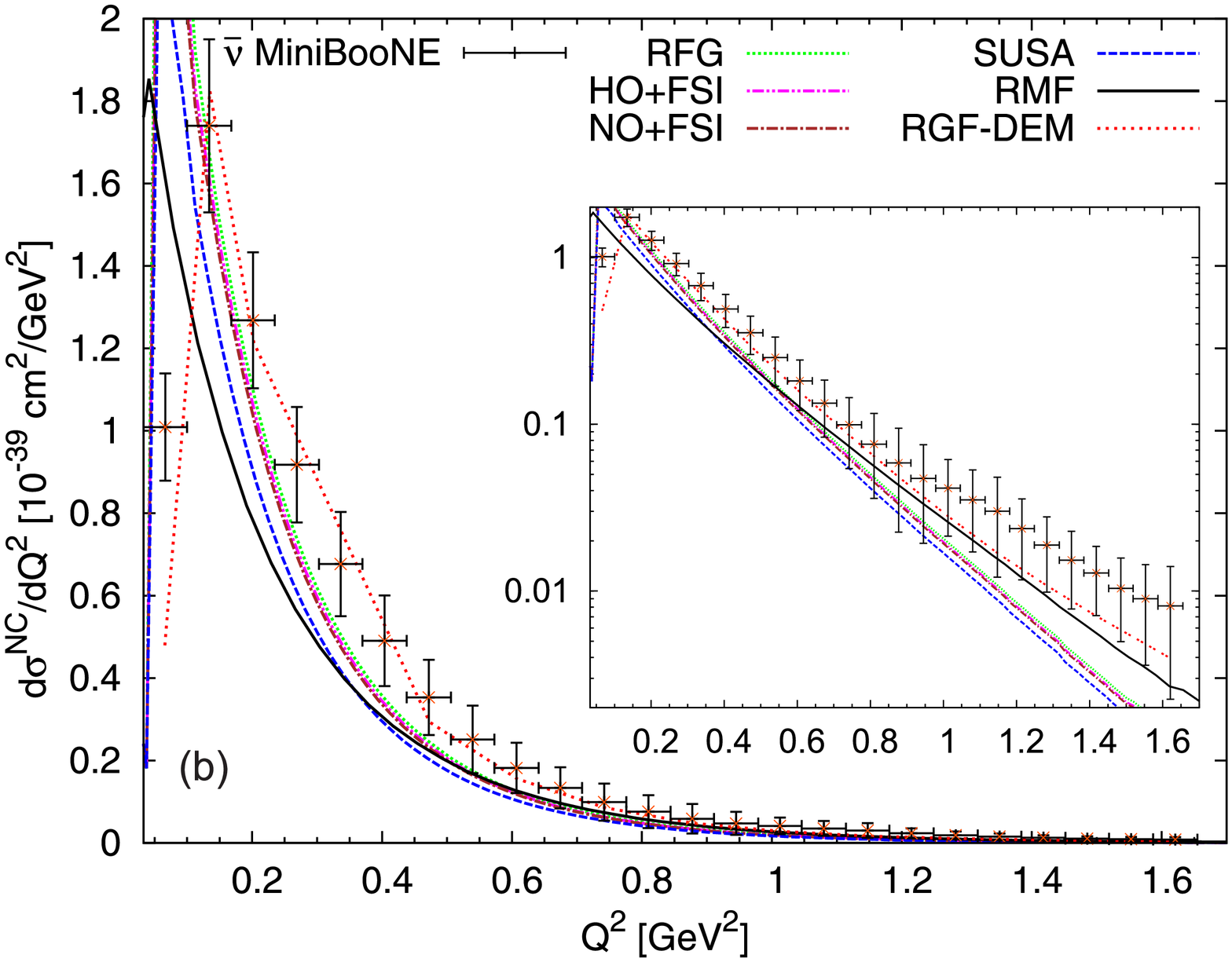}
\caption{(Color online) NCQE neutrino [panel (a), $\nu N\to\nu N$] and antineutrino [panel (b), $\overline{\nu}N\to\overline{\nu}N$] flux-averaged differential cross section computed using the RFG, HO+FSI, NO+FSI, SUSA scaling functions, RGF and RMF models and compared with the MiniBooNE data~\cite{ncsf42, ncsf45}. The results correspond to the world-average axial mass $M_A = 1.032$~GeV and strangeness $\Delta s = 0$. The error bars do not account for the normalization uncertainty of 18.1\% (19.5\%) in the $\nu$($\overline{\nu}$) case.}\label{fig01}
\end{figure}

\begin{figure}[t!]
\includegraphics[width=0.95\columnwidth]{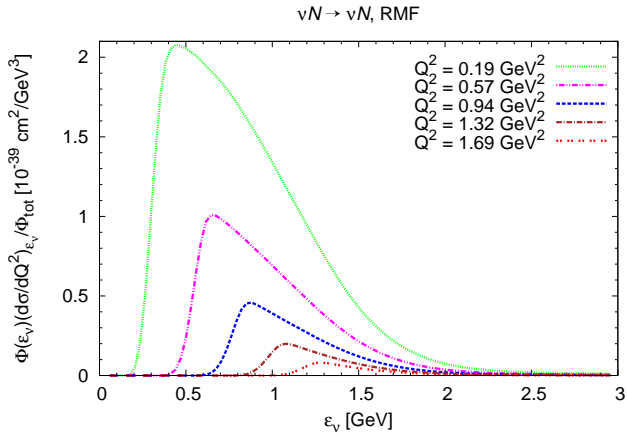}
\caption{(Color online) Differential cross section $d\sigma/dQ^2$ for $\nu$NCQE scattering per bound nucleon in the $^{12}$C nucleus calculated within RMF model and multiplied by the MiniBooNE flux. The results are given at several fixed values of four-momentum transfer squared ($Q^2$).}\label{fig01a}
\end{figure}

\begin{figure}[t]
\includegraphics[width=0.95\columnwidth]{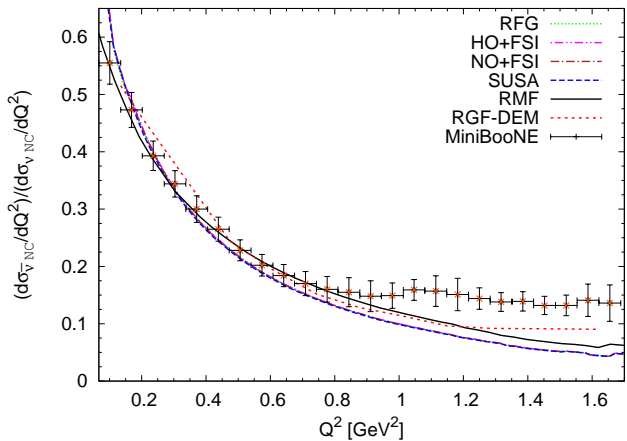}
\caption{(Color online) Ratio of the antineutrino to neutrino NCQE scattering cross section computed using the RFG, HO+FSI, NO+FSI, SUSA, RGF, and RMF approaches and compared with the MiniBooNE data~\cite{ncsf45}.}\label{fig04}
\end{figure}

\begin{figure*}[t]
\centering\includegraphics[width=160mm,angle=0]{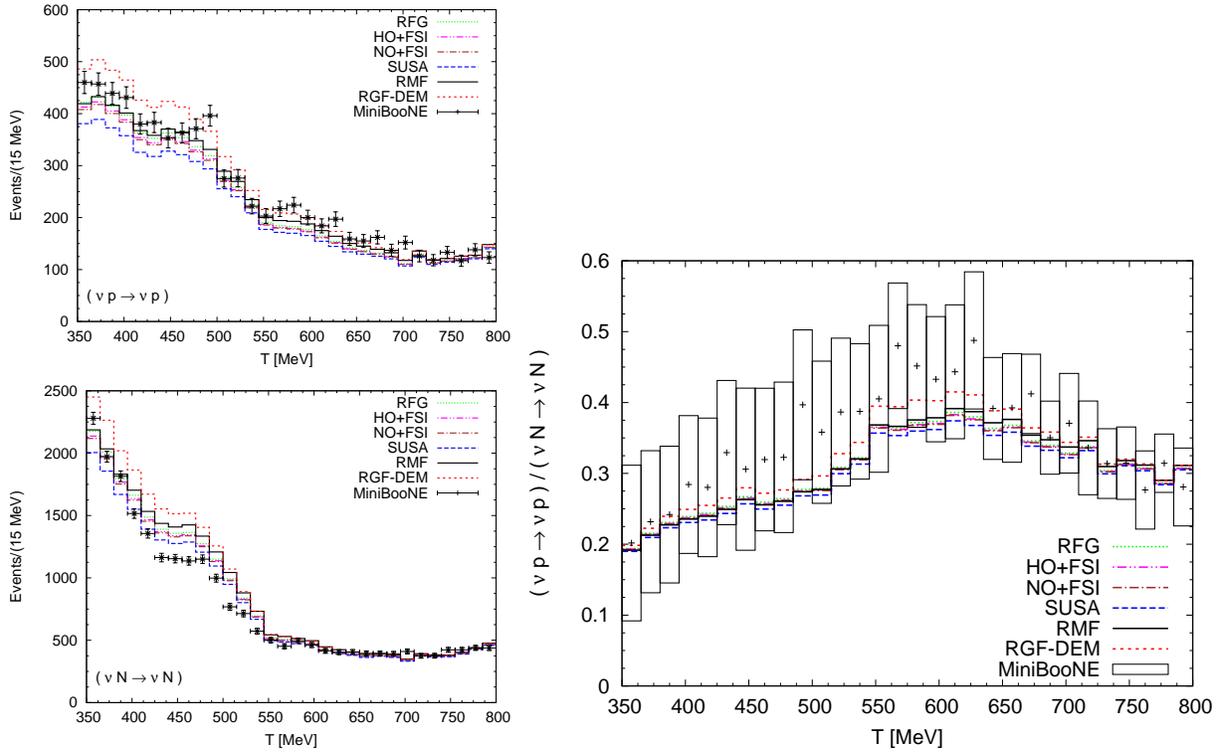}
\caption{(Color online) The RFG, HO+FSI, NO+FSI, SUSA, RGF, and RMF predictions, after the folding procedure, compared with the histograms of the numerator (top-left panel) and denominator (bottom-left panel) of the ratio. The error bars in the left panel represent only the statistical uncertainty, computed as the square root of the event number. The corresponding ratio is shown in the right panel of the figure. The axial mass and strangeness have been assumed to be the standard axial mass value and zero strangeness. Data are taken from~\cite{ncsf42, ncsf68, ncsf80}.}\label{fig03}
\end{figure*}

\begin{figure*}[t]
\includegraphics[width=0.95\columnwidth]{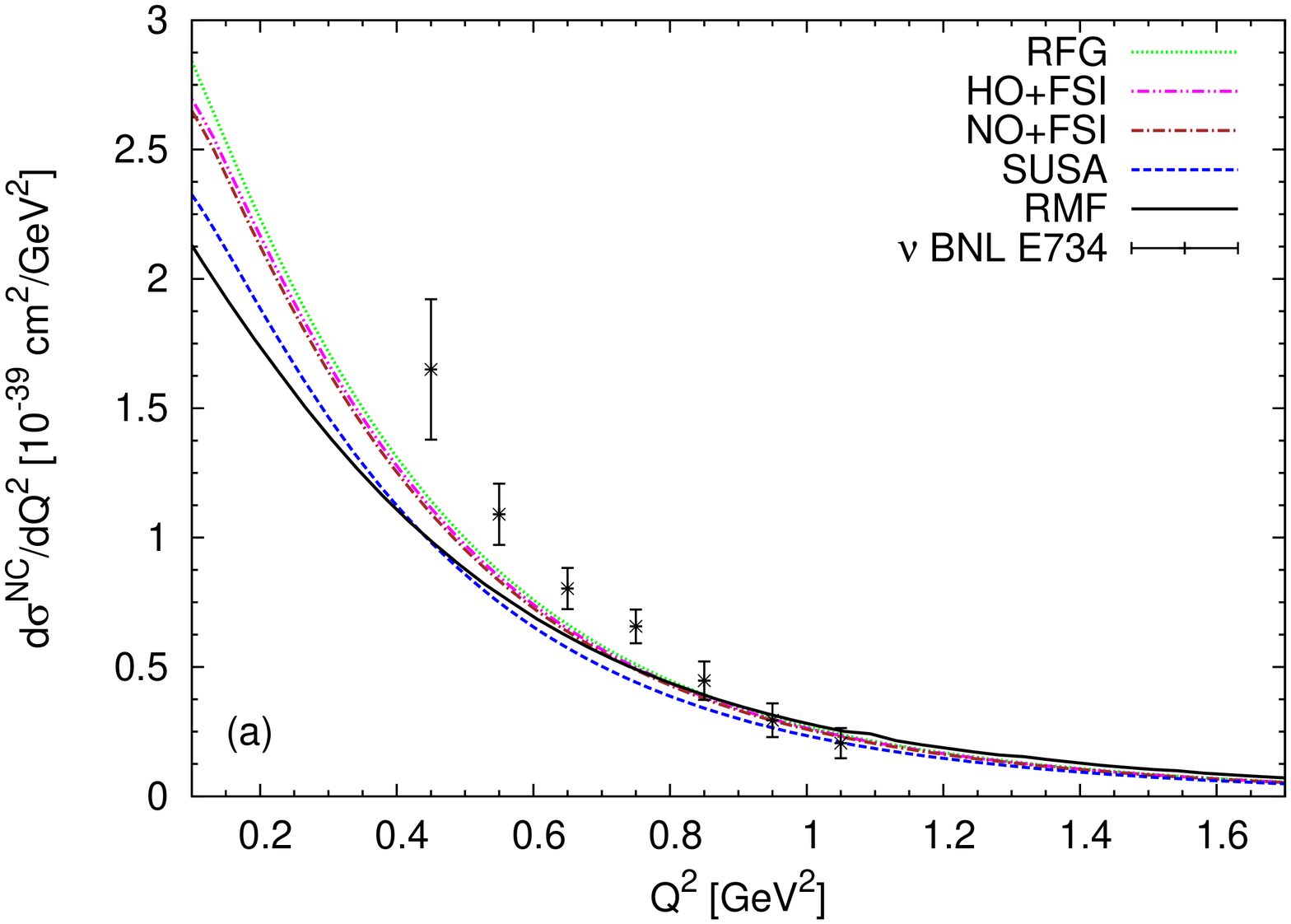}\hfill%\\[5pt]
\includegraphics[width=0.95\columnwidth]{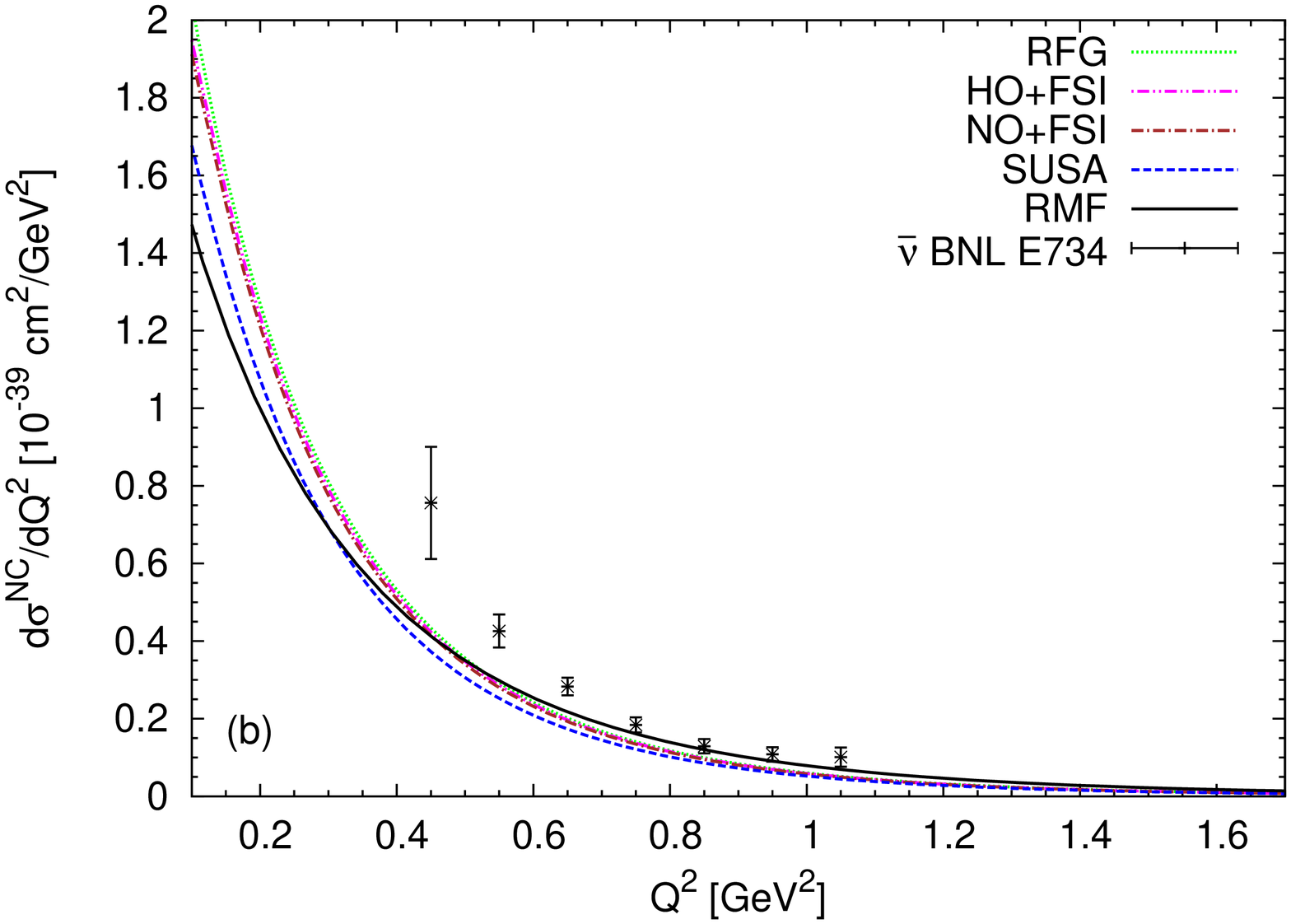}
\caption{(Color online) NCQE flux-averaged cross section: $\nu p \to \nu p$ [panel (a)] and $\overline{\nu}p \to \overline{\nu} p$ [panel (b)] compared with the BNL E734 experimental data~\cite{ncsf60}. Our results are evaluated using the RFG, HO+FSI, NO+FSI, SUSA scaling functions, and RMF model with the standard value of the axial-vector dipole mass $M_A = 1.032$~GeV$^2$ and strangeness $\Delta s = 0$. The error bars do not include the normalization uncertainty of 11.2\% (10.4\%) in the $\nu$($\overline{\nu}$) case.}\label{fig02}
\end{figure*}

In this section the theoretical predictions of the  RFG, HO+FSI, NO+FSI, SUSA scaling functions are compared with the data measured by the MiniBooNE and BNL Collaborations. The comparison is performed also with the results of the RMF and RGF models, which are based on the same relativistic mean-field model for nuclear structure but on a different treatment of FSI. In the RMF model FSI are described by the same relativistic mean field potential describing the initial nucleon state;  the description of FSI in the RGF is based on the use of a complex optical  potential. Details of the RGF model can be found, for instance, in~\cite{RGFa,RGFb}. The results of the RMF and RGF models have been already and widely compared in~\cite{compee} for the inclusive QE electron scattering, in~\cite{ncsf54,compcc} for CCQE and in~\cite{ncsf54c} for NCQE neutrino scattering. The RGF calculations presented in this work have been carried out with the recent democratic optical potential of~\cite{ROP} (RGF-DEM).

The comparison between theory and experiment for the NCQE flux-averaged MiniBooNE (anti)neutrino cross section is presented in Fig.~\ref{fig01}. Here we compare the predictions using the RFG, HO+FSI, NO+FSI, SUSA scaling functions, and RMF model with the data. As usual in NC reactions, in this work, the variable $Q^2$ is defined as $Q^2=2 M_N T_N$, where $M_N$ and $T_N$ are the mass and kinetic energy of the outgoing nucleon, respectively. In order to compare with MiniBooNE we evaluate the following differential cross section per nucleon:
\begin{align}\label{trueCS}
\dfrac{d\sigma_{\nu(\overline{\nu})N\to\nu(\overline{\nu})N}}{dQ^2} =&
\dfrac{1}{7}  C_{\nu(\overline{\nu}) p,\text{H}}(Q^2)
\dfrac{d\sigma_{\nu(\overline{\nu}) p\to \nu(\overline{\nu}) p, \text{H}} }{dQ^2}\notag\\
+&\dfrac{3}{7}  C_{\nu(\overline{\nu}) p,\text{C}}(Q^2)
\dfrac{d\sigma_{\nu(\overline{\nu}) p\to \nu(\overline{\nu}) p, \text{C}} }{dQ^2}\notag\\
+&\dfrac{3}{7}  C_{\nu(\overline{\nu}) n,\text{C}}(Q^2)
\dfrac{d\sigma_{\nu(\overline{\nu}) n\to \nu(\overline{\nu}) n, \text{C}}}{dQ^2},
\end{align}
where $C_{\nu(\overline{\nu}) p,\text{H}}(Q^2)$, $C_{\nu(\overline{\nu}) p,\text{C}}(Q^2)$, and $C_{\nu(\overline{\nu}) n,\text{C}}(Q^2)$ are the efficiency correction functions, given in Refs.~\cite{ncsf45, ncsf42}, of three different processes: the (anti)neutrino scattering off free protons in the hydrogen atom, the bound protons in the carbon atom and the bound neutrons in the carbon atom. ${d\sigma_{\nu(\overline{\nu}) p\to \nu(\overline{\nu}) p, \text{H}} }/{dQ^2}$, ${d\sigma_{\nu(\overline{\nu}) p\to \nu(\overline{\nu}) p, \text{C}} }/{dQ^2}$, and ${d\sigma_{\nu(\overline{\nu}) n\to \nu(\overline{\nu}) n, \text{C}}}/{dQ^2}$ are the theoretical $\nu$($\overline{\nu}$)NCQE cross sections on free protons (per free proton), on bound protons (per bound proton), and on bound neutrons (per bound neutron). As can be seen from Fig.~\ref{fig01} the theoretical results corresponding to all models except the RGF-DEM underestimate the neutrino data in the region between $0.1<Q^2<0.7$~GeV$^2$, while all theories are within the error bars for higher $Q^2$. On the other hand the same models underestimate the antineutrino data at high $Q^2$. This is clearly seen in the insets of Fig.~\ref{fig01}, where the cross sections are represented in logarithmic scale. The RGF-DEM  results are larger than the results of the other models and in generally good agreement with the data over the entire  $Q^2$ region considered in the figure. The enhancement  of the RGF cross sections is due to the contribution of final-state channels that are recovered by the imaginary part of the optical potential and that are not included in the other models.

In order to better understand the behaviour of our results with $Q^2$, we present in Fig.~\ref{fig01a} the differential cross sections $d\sigma/dQ^2$ for $\nu$NCQE scattering (per bound nucleon in the $^{12}$C nucleus) calculated within the RMF model and multiplied by the MiniBooNE flux, at several fixed values of $Q^2$. $\Phi_\text{tot}$ is the total integrated $\nu_\mu$ flux factor for the MiniBooNE experiment:
\begin{equation}
\Phi_\text{tot}=\int\Phi(\varepsilon_\nu)d\varepsilon_\nu.\label{Eq14}
\end{equation}
As can be seen from Fig.~\ref{fig01a} for small values of $Q^2$ the contribution to the cross sections is accumulated from a large neutrino energy range (for example, at $Q^2=0.19$~GeV$^2$ the contributions to the cross section come from neutrino energies from $0.2$ to $2.5$~GeV), whereas for higher $Q^2$ values, the energy range which gives contribution to the cross section becomes smaller (for example, at $Q^2=1.69$~GeV$^2$ contributions to the cross section come from neutrino energies from $1.1$ to $2.5$~GeV). This suggests that the discrepancy between the RMF result and the experimental data at low $Q^2$ can be ascribed to the failure of the impulse approximation for low (anti)neutrino energies. For the other models (leaving apart the RGF) the situation is different for $\nu$ and $\bar\nu$ scattering. To better illustrate this difference we show in Fig.~\ref{fig04} our predictions for the ratio of the antineutrino to neutrino NCQE scattering cross sections. The results are compared with the recently published MiniBooNE data~\cite{ncsf45}. Since both measurements were made in the same beamline and with the same detector, it is expected that the bin-to-bin ratio of the two cross section measurements would cancel the common systematic errors. As can be seen from Fig.~\ref{fig04} the results of all models are in agreement with the data up to $Q^2<0.8$~GeV$^2$ (actually the RMF and RGF results are in good agreement up to $Q^2\sim1$~GeV$^2$), whereas for bigger $Q^2$ all models underpredict the data. This is so because the theoretical antineutrino NCQE cross section underpredicts the data at high $Q^2$ (see Fig.~\ref{fig01}). It has to be noted that the results using different scaling functions almost coincide, since the cross sections are proportional to s.n. cross sections multiplied by the scaling function [see Eq.~(\ref{Eq01})] and the ratio is proportional to the ratio of the s.n. antineutrino to s.n. neutrino cross sections. The updated version of the SuSA model, ``SuSAv2'', gives an interesting possibility in the studies of the NC cross sections and particularly of the mentioned ratio. This is related to the use of different parametrizations of the transverse and longitudinal scaling functions in SuSAv2. Work along this line is in progress.

For completeness we also present in Fig.~\ref{fig03} the spectra corresponding to the numerator and denominator entering the ratio between $\nu$-scattering from proton and nucleon (proton plus neutron) (left panels) and the ratio computed by dividing of the two samples (right panels) within the various models and compared with the MiniBooNE data~\cite{ncsf42}. The numerator and denominator data (left panels of Fig.~\ref{fig03}) are taken from~\cite{ncsf80}, where the data are reported without the corresponding errors (so, in the figure only statistical errors are included). In the calculations the axial mass and strangeness have been assumed to be the standard axial mass and zero strangeness. We note that the dispersion between the models tends to cancel when this ratio is considered. This result clearly shows that the proton/nucleon ratio is very insensitive to nuclear model effects and to final state interactions, and hence, it may provide information that improves our present knowledge on the electroweak nucleon structure, in particular, the nucleon's strangeness. In particular, for kinetic energies of the outgoing nucleon $T_N > 350$~MeV our models give results which are in good agreement with the experimental data (left panels of Fig.~\ref{fig03}), while for the ratio our theoretical results are within the error bars at all kinematics (right panel of Fig.~\ref{fig03}).

We now compare the results obtained with our models with the BNL E734 experimental data. The mean value of neutrino (antineutrino) energy is $1.3$~GeV ($1.2$~GeV) for BNL experiment, while for MiniBooNE experiment it is $788$~MeV ($665$~MeV). In Fig.~\ref{fig02} the differential cross sections evaluated using the RFG, HO+FSI, NO+FSI, SUSA scaling functions, and RMF model are compared with NCQE $\nu p \to \nu p$ [panel (a)] and $\overline{\nu}p \to \overline{\nu} p$ [panel (b)] BNL E734 experimental data. The BNL E734 experiment was performed using a 170-metric-ton high-resolution target-detector in a horn-focused (anti)neutrino at the Brookhaven National Laboratory. The cross sections results show a behaviour similar to those of the MiniBooNE experiment. The latter (using the Cherenkov detector filled with mineral oil (CH$_2$)) is sensitive to both $\nu(\overline{\nu})p$ and $\nu(\overline{\nu})n$ NCQE scattering~\cite{ncsf42, ncsf45}. It has been known for some time (see, \emph{e.g.},~\cite{ncsf70, Alberico:1998qw, ncsf79}) that the $\Delta s$-dependence of the NCQE neutrino-nucleon cross section is very mild. This results from a cancellation between the effect of $\Delta s$ on the proton and neutron contributions, which are affected differently by the axial strangeness: by changing $\Delta s$ from zero to a negative value the proton cross section gets enhanced while the neutron one is reduced, so that the net effect on the total cross section is very small. NCQE $\nu(\overline{\nu})p$ differential cross sections were measured in the BNL E734 experiment, which are sensitive to the values of $\Delta s$ (there is not a cancellation effect). The BNL E734 experimental data can be reproduced within our models in principle by the fit of the axial strangeness without change of the axial mass value.

Here we would like to mention that, first, our calculations using NO and HO single-particle wave functions in $n_i(p)$ in the spectral function Eq.~(\ref{HF+lorent}) with FSI and without FSI  show that the inclusion of FSI effects leads to a small change (a depletion) of the cross sections. Second, the results for the cross sections obtained using realistic spectral function $S(p,E)$ with single-particle momentum distributions $n_i(p)$ (see Eq.~(\ref{HF+lorent})) that include Jastrow short-range NN correlations (accounted for in the NO's) can be compared in Figs.~\ref{fig01},~\ref{fig04},~\ref{fig03}, and~\ref{fig02} with those when NN correlation are not included (RFG and HO). It can be seen that, similarly to the case of CCQE neutrino scattering (see Ref.~\cite{ncsf67}), the differences between results in correlated and non-correlated approaches are small, thus showing that the process is not too sensitive to the specific treatment of the bound state.

\section{Conclusions \label{sec:4}}

This work complements previous studies focused on charged-current quasielastic (CCQE) neutrino-nucleus scattering processes that were performed making use of a realistic spectral function. Here we extend this analysis to the case of neutral-current (NC) neutrino processes, and compare our theoretical predictions with data measured by MiniBooNe and BNL Collaborations. Contrary to CCQE reactions, where the final lepton is detected, in the NC case one has no information on the energy and momentum of the ejected neutrino. Hence the transferred four-momentum cannot be determined. This, as already discussed in some previous works, makes the description of the reaction  mechanism not so clear and some caution should be also drawn on the ``validity'' of scaling arguments when applied to NC. However, our previous studies give us confidence in the reliability of our calculations and their application to describe  the present experimental data measured at different facilities.

The main objective of this work is centered in the use of a realistic spectral function, that accounts for short-range NN correlations, and has also a realistic energy dependence. This function gives a scaling function in accordance with electron scattering data and it can be used for a wide range of neutrino energies. Therefore, the use of this spectral function to describe the general reaction mechanism involved in NC neutrino-nucleus scattering processes can provide very valuable information that can be confronted with results obtained with other theoretical approaches. In this sense, we compare our spectral function-based predictions with the results provided by the SuSA, RMF and RGF models largely used by us in the past. The discrepancies found can help disentangling effects directly linked to particular ingredients in the process: final state interactions, nucleon correlations, effects beyond the impulse approximation, \emph{etc.}

The predictions of our model agree in general with previous results although with some peculiarities that should be emphasized. Our calculations showed that the inclusion of FSI effects in the spectral function-based calculations leads to a slight depletion of the cross section being in close agreement with the RFG prediction. The inclusion of FSI effects in the RGF model leads to larger cross sections, in good agreement with the data. On the contrary, SuSA and, in particular, RMF approaches lead to significantly smaller differential cross sections at low values of $|Q^2|$ ($\leq 0.6$--$0.8$~GeV$^2$) departing also from the data. This behavior can be seen for the two experiments considered in the work: MiniBooNe and BNL. Another point of relevance when comparing the different models is the softer $Q^2$ dependence (with a smaller slope) shown by the RMF cross section. Whereas it is clearly below the other curves at low $Q^2$ (up to $\approx 0.5$--$0.6$~GeV$^2$), it crosses them providing the largest contribution at higher $Q^2$.  This result can be taken as an indication of the particular sensitivity of NC processes to the specific description of FSI effects. It may be also connected with the increasing tail in the scaling function provided by the RMF model at larger $Q^2$-values. This is a consequence of the enhancement of the lower components in the relativistic nucleon wave functions, particularly, for the final state.

All our calculations are based on the impulse approximation, \emph{i.e.}, they do not include effects beyond the one-body approach, for example $2p-2h$ contributions induced by meson exchange currents. These ingredients have been shown to be very important in the analysis of neutrino-nucleus scattering processes. In particular, they produce a significant enhancement in the cross section at low-to-moderate values of the transferred four momentum. This is consistent with our predictions that clearly underestimate data for such kinematical regions. On the contrary, the accordance improves at higher $Q^2$. This is strictly true for neutrinos at MiniBooNE. In the case of antineutrinos, MiniBooNE data at $Q^2\geq 1$--$1.2$~GeV$^2$ are higher than theoretical predictions, the RMF results being closer to the experiment. This behavior leads also to the significant discrepancy observed for the antineutrino/neutrino ratio (Fig.~\ref{fig04}). More studies are needed in order to understand these differences at medium-large $Q^2$-values. This could be related to a different role played by $2p-2h$ contributions and MEC for neutrinos and antineutrinos. Work is in progress to evaluate the impact of $2p-2h$ excitations on these results.

\section*{Acknowledgments}

This work was partially supported by Spanish DGI and FEDER funds (FIS2011-28738-C02-01, FPA2013-41267), by the Junta de Andalucia, by the Spanish Consolider-Ingenio 2000 program CPAN (CSD2007-00042), by the Campus of Excellence International (CEI) of Moncloa project (Madrid) and Andalucia Tech, by the Istituto Nazionale di Fisica Nucleare under Contract MANYBODY, as well as by the Bulgarian National Science Fund under contracts No. DFNI-T02/19 and DFNI-E02/6. M.V.I. is grateful for the warm hospitality given by the UCM and for financial support during his stay there from the SiNuRSE action within the ENSAR European project. A.N.A. acknowledges financial support from the Universidad de Sevilla under the Program: ``IV Plan Propio de Investigaci\'on. Movilidad de Investigadores''. R.G.J. acknowledges financial help from VPPI-US (Universidad de Sevilla).


\begin{thebibliography}{99}

\bibitem{ncsf01} G. B. West, Phys. Rep. {\bf 18}, 263 (1975).

\bibitem{ncsf02} I. Sick, D. B. Day, and J. S. McCarthy, Phys. Rev. Lett. {\bf 45}, 871 (1980).

\bibitem{ncsf03} C. Ciofi degli Atti, E. Pace, and G. Salm\`{e}, Phys. Rev. C {\bf 36}, 1208 (1987).

\bibitem{ncsf04} D. B. Day, J. S. McCarthy, T. W. Donnelly, and I. Sick, Annu. Rev. Nucl. Part. Sci. {\bf 40}, 357 (1990).

\bibitem{ncsf05} C. Ciofi degli Atti, E. Pace, and G. Salm\`{e}, Phys. Rev. C {\bf 43}, 1155 (1991).

\bibitem{ncsf06} C. Ciofi degli Atti and S. Simula, Phys. Rev. C {\bf 53}, 1689 (1996).

\bibitem{ncsf07} C. Ciofi degli Atti and G. B. West, arXiv:nucl-th/9702009.

\bibitem{ncsf08} C. Ciofi degli Atti and G. B. West, Phys. Lett. B \textbf{458}, 447 (1999).

\bibitem{ncsf09} D. Faralli, C. Ciofi degli Atti, and G. B. West, in \emph{Proceedings of 2nd International Conference on Perspectives in Hadronic Physics}, ICTP, Trieste, Italy, 1999, edited by S. Boffi, C. Ciofi degli Atti, and M. M. Giannini (World Scientific, Singapore, 2000), p.~75.

\bibitem{ncsf10} W. M. Alberico, A. Molinari, T. W. Donnelly, E. L. Kronenberg, and J. W. Van Orden, Phys. Rev. C \textbf{38}, 1801 (1988).

\bibitem{ncsf11} M. B. Barbaro, R. Cenni, A. De Pace, T. W. Donnelly, and A. Molinari, Nucl. Phys. A \textbf{643}, 137 (1998).

\bibitem{ncsf12} T. W. Donnelly and I. Sick, Phys. Rev. Lett. \textbf{82}, 3212 (1999).

\bibitem{ncsf13} T. W. Donnelly and I. Sick, Phys. Rev. C \textbf{60}, 065502 (1999).

\bibitem{ncsf14} C. Maieron, T. W. Donnelly, and I. Sick, Phys. Rev. C \textbf{65}, 025502 (2002).

\bibitem{ncsf15} M.~B.~Barbaro, J.~A.~Caballero, T.~W.~Donnelly and C.~Maieron, Phys. Rev. C {\bf 69}, 035502 (2004).

\bibitem{ncsf16} A. N. Antonov, M. K. Gaidarov, D. N. Kadrev, M. V. Ivanov, E. Moya de Guerra, and J. M. Udias, Phys. Rev. C \textbf{69}, 044321 (2004).

\bibitem{ncsf17} A. N. Antonov, M. K. Gaidarov, M. V. Ivanov, D. N. Kadrev, E. Moya de Guerra, P. Sarriguren, and J. M. Udias, Phys. Rev. C \textbf{71}, 014317 (2005).

\bibitem{ncsf18} A. N. Antonov, M. V. Ivanov, M. K. Gaidarov, E. Moya de Guerra, P. Sarriguren, and J. M. Udias, Phys. Rev. C \textbf{73}, 047302 (2006).

\bibitem{ncsf19} A. N. Antonov, M. V. Ivanov, M. K. Gaidarov, E. Moya de Guerra, J. A. Caballero, M. B. Barbaro, J. M. Udias, and P. Sarriguren, Phys. Rev. C \textbf{74}, 054603 (2006).

\bibitem{ncsf20} A. N. Antonov, M. V. Ivanov, M. K. Gaidarov, and E. Moya de Guerra, Phys. Rev. C \textbf{75}, 034319 (2007).

\bibitem{ncsf21} O. Benhar, D. Day, and I. Sick, Rev. Mod. Phys. \textbf{80}, 189 (2008).

\bibitem{ncsf22} Y. Fukuda {\it et al.} (The Super-Kamiokande Collaboration), Phys. Rev. Lett. {\bf 81}, 1562 (1998); M. H. Ahn {\it et al.} (K2K Collaboration), {\it ibid.} {\bf 90}, 041801 (2003); Q.-R. Ahmad {\it et al.} (SNO Collaboration), {\it ibid.} {\bf 87}, 071301 (2001); {\bf 89}, 011301 (2002); K. Eguchi {\it et al.} (KamLAND Collaboration), {\it ibid.} {\bf 90}, 021802 (2003); C. Athanassopoulos {\it et al.} (LSND Collaboration), {\it ibid.} {\bf 77}, 3082 (1996); {\bf 81}, 1774 (1998).

\bibitem{ncsf23} A.~N.~Antonov, V.~A.~Nikolaev, and I.~Zh.~Petkov, Bulg. J. Phys. \textbf{6}, 151 (1979); Z. Phys. A \textbf{297}, 257 (1980); \textit{ibid.} \textbf{304}, 239 (1982); Nuovo Cimento A \textbf{86}, 23 (1985); Nuovo Cimento A \textbf{102}, 1701 (1989); A.~N.~Antonov, D.~N.~Kadrev, and P.~E.~Hodgson, Phys. Rev. C \textbf{50}, 164 (1994).

\bibitem{ncsf24} A.~N.~Antonov, P.~E.~Hodgson, and I.~Zh.~Petkov, \textit{Nucleon Momentum and Density Distributions in Nuclei} (Clarendon Press, Oxford, 1988); \textit{Nucleon Correlations in Nuclei} (Springer-Verlag, Berlin-Heidelberg-New York, 1993).

\bibitem{ncsf25} A.~N.~Antonov, M.~V.~Ivanov, M.~B.~Barbaro, J.~A.~Caballero, E.~Moya de Guerra, M.~K.~Gaidarov, Phys. Rev. C \textbf{75}, 064617 (2007).

\bibitem{ncsf26} J. E. Amaro, M. B. Barbaro, J. A. Caballero, T. W. Donnelly, A. Molinari, and I. Sick, Phys. Rev. C \textbf{71}, 015501 (2005).

\bibitem{ncsf27} J. A. Caballero, J.~E.~Amaro, M. B. Barbaro, T. W. Donnelly, C. Maieron, and J. M. Udias, Phys. Rev. Lett. \textbf{95}, 252502 (2005).

\bibitem{ncsf28} J.A. Caballero, Phys. Rev. C \textbf{74}, 015502 (2006).

\bibitem{ncsf29} J.~E.~Amaro, M.~B.~Barbaro, J.~A.~Caballero, T.~W.~Donnelly, and C.~Maieron, Phys. Rev. C {\bf 71}, 065501 (2005).

\bibitem{ncsf30} J.~E.~Amaro, M.~B.~Barbaro, J.~A.~Caballero, and T.~W.~Donnelly, Phys. Rev. C {\bf 73}, 035503 (2006).

\bibitem{ncsf31} A. Meucci, C. Giusti, and F. D. Pacati, Nucl. Phys. A \textbf{744}, 307 (2004).

\bibitem{ncsf32} M. B. Barbaro, Nucl. Phys. B, Proc. Suppl. \textbf{159}, 186 (2006).

\bibitem{ncsf33} M. C. Martinez, P. Lava, N. Jachowicz, J. Ryckebusch, and J. M. Udias, Phys. Rev. C \textbf{73}, 024607 (2006).

\bibitem{ncsf34} J. Nieves, M. Valverde, and M. J. Vicente-Vacas, Nucl. Phys. B, Proc. Suppl. \textbf{155}, 263 (2006).

\bibitem{ncsf35} M. B. Barbaro, J. E. Amaro, J. A. Caballero, T. W. Donnelly, and A. Molinari, Nucl. Phys. B, Proc. Suppl. \textbf{155}, 257 (2006).

\bibitem{ncsf36} C. Maieron, M. C. Martinez, J. A. Caballero, and J. M. Udias, Phys. Rev. C \textbf{68}, 048501 (2003).

\bibitem{ncsf37} A. Meucci, C. Giusti, and F. D. Pacati, Nucl. Phys. A \textbf{739}, 277 (2004); Nucl. Phys. A \textbf{773}, 250 (2006).

\bibitem{ncsf38} O. Benhar, Nucl. Phys. B, Proc. Suppl. \textbf{139}, 15 (2005); O. Benhar and N. Farina, Nucl. Phys. B, Proc. Suppl. \textbf{139}, 230 (2005); O. Benhar, N. Farina, H. Nakamura, M. Sakuda, and R. Seki, Nucl. Phys. B, Proc. Suppl. \textbf{155}, 254 (2006); Phys. Rev. D \textbf{72}, 053005 (2005).

\bibitem{ncsf39} G. Co', Nucl. Phys. B, Proc. Suppl. \textbf{159}, 192 (2006); A. Botrugno and G. Co', Nucl. Phys. A \textbf{761}, 200 (2005); M. Martini, G. Co', M. Anguiano, and A. M. Lallena, Phys. Rev. C \textbf{75}, 034604 (2007).

\bibitem{ncsf40} T. Leitner, L. Alvarez-Ruso, and U. Mosel, Phys. Rev. C \textbf{73}, 065502 (2006).

\bibitem{ncsf41} B. Szczerbinska, T. Sato, K. Kubodera, and T.-S. H. Lee, Phys. Lett. B \textbf{649}, 132 (2007).

%\cite{AguilarArevalo:2010zc}
\bibitem{AguilarArevalo:2010zc}
  A. A. Aguilar-Arevalo {\emph et al.}  (MiniBooNE Collaboration),
  %``First Measurement of the Muon Neutrino Charged Current Quasielastic Double Differential Cross Section,''
  Phys. Rev. D \textbf{81}, 092005 (2010).
%  [arXiv:1002.2680 [hep-ex]].
  %%CITATION = ARXIV:1002.2680;%%
  %231 citations counted in INSPIRE as of 28 Nov 2014

\bibitem{ncsf42} A. A. Aguilar-Arevalo \emph{et al.} (MiniBooNE Collaboration), Phys. Rev. D \textbf{82}, 092005 (2010).

\bibitem{ncsf60} K. Abe \emph{et al.}, Phys. Rev. Lett. \textbf{56}, 1107 (1986); \textbf{56}, 1883(E) (1986); L. A. Ahrens \emph{et al.}, Phys. Rev. D \textbf{35}, 785 (1987).

\bibitem{ncsf44} A.~A.~Aguilar-Arevalo \emph{et al.} [MiniBooNE Collaboration], Phys. Rev. D \textbf{88}, 032001 (2013). %arXiv:1301.7067 [hep-ex].

\bibitem{ncsf45} A.~A.~Aguilar-Arevalo {\it et al.}  [MiniBooNE Collaboration], Phys. Rev. D \textbf{91}, 012004 (2015). %arXiv:1309.7257 [hep-ex].

\bibitem{ncsf46} A.~M.~Ankowski, Phys. Rev. C {\bf 86}, 024616 (2012).

\bibitem{ncsf47} P. Lipari, Nucl. Phys. B, Proc. Suppl. \textbf{112}, 274 (2002); G. P. Zeller, arXiv:hep-ex/0312061.

\bibitem{ncsf48} A.~Meucci, C.~Giusti and F.~D.~Pacati, Phys.\ Rev.\ D {\bf 84}, 113003 (2011).

\bibitem{ncsf49}  A.~V.~Butkevich and D.~Perevalov, Phys.\ Rev.\ C {\bf 84}, 015501 (2011).

\bibitem{ncsf50} O.~Benhar and G.~Veneziano, Phys.\ Lett.\ B {\bf 702}, 433 (2011).

\bibitem{ncsf51} M.~Martini, M.~Ericson and G.~Chanfray, Phys.\ Rev.\ C {\bf 84}, 055502 (2011).

\bibitem{ncsf52} J.~Nieves, I.~R.~Simo and M.~J.~V.~Vacas, Phys.\ Lett.\ B {\bf 707}, 72 (2012).

\bibitem{ncsf53} J.~E.~Amaro, M.~B.~Barbaro, J.~A.~Caballero, T.~W.~Donnelly and J.~M.~Udias, Phys.\ Rev.\ D {\bf 84}, 033004 (2011).

\bibitem{ncsf54} A.~Meucci, M.~B.~Barbaro, J.~A.~Caballero, C.~Giusti and J.~M.~Udias, Phys.\ Rev.\ Lett.\  {\bf 107}, 172501 (2011).

\bibitem{ncsf55} V.~Bernard, L.~Elouadrhiri and Ulf-G~Mei{\ss}ner, J.\ Phys.\ G {\bf 28}, R1 (2002).

\bibitem{ncsf56} M.~Martini, M.~Ericson, G.~Chanfray and J.~Marteau, Phys.\ Rev.\  C {\bf 80}, 065501 (2009).

\bibitem{ncsf57} M.~Martini, M.~Ericson, G.~Chanfray, and J.~Marteau, Phys. Rev. C {\bf81}, 045502 (2010).

\bibitem{raul1} R.~Gonz\'{a}lez-Jim\'{e}nez, M.~V.~Ivanov, M.~B.~Barbaro, J.~A.~Caballero, J.~M.~Udias, Phys. Lett. B \textbf{718}, 1471 (2013).

\bibitem{raul2} M.~V.~Ivanov, R.~Gonz\'{a}lez-Jim\'{e}nez, J.~A.~Caballero, M.~B.~Barbaro, T.~W.~Donnelly, J.~M.~Udias, Phys. Lett. B \textbf{727}, 265 (2013).

\bibitem{ncsf54c} R.~Gonzalez-Jimenez, J.~A.~Caballero, A.~Meucci, C.~Giusti, M.~B.~Barbaro, M.~I.~Ivanov and J.~M.~Udias, Phys.\ Rev.\ C  {\bf 88}, 025502 (2013).

\bibitem{ncsf62} A. Meucci and C. Giusti, Phys. Rev. D \textbf{89}, 057302 (2014).

\bibitem{compee} A.~Meucci, J.~A.~Caballero, C.~Giusti, F.~D.~Pacati and J.~M.~Udias, Phys. Rev. C {\bf 80}, 024605 (2009).

\bibitem{ncsf54a} A.~Meucci and C.~Giusti, Phys.\ Rev.\ D\  {\bf 85}, 093002 (2012).

%\bibitem{ncsf54b} A.~Meucci, C.~Giusti and F.~D.~Pacati, Phys.\ Rev.\ D\  {\bf 84}, 113003 (2011).

\bibitem{ncsf58} J.~E.~Amaro, M.~B.~Barbaro, J.~A.~Caballero, T.~W.~Donnelly and C.~F.~Williamson, Phys.\ Lett.\  B {\bf 696}, 151 (2011).

\bibitem{Gonzalez-Jimenez:2014eqa}
  R.~Gonz\'{a}lez-Jim\'{e}nez, G.~D.~Megias, M.~B.~Barbaro, J.~A.~Caballero and T.~W.~Donnelly, Phys.\ Rev.\ C {\bf 90}, 035501 (2014).

\bibitem{ncsf59} O. Lalakulich, K. Gallmeister, and U. Mosel, Phys. Rev. C \textbf{86}, 014614 (2012).

\bibitem{ncsf43} V.~Lyubushkin {\it et al.}  [NOMAD Collaboration],  Eur.\ Phys.\ J.\ C {\bf 63}, 355 (2009).

%L. A. Ahrens \emph{et al.}, Phys. Rev. D \textbf{35}, 785 (1987); G.~T.~Garvey, W.~C.~Louis, and D.~H.~White, Phys. Rev. C \textbf{48}, 761 (1993); L. Bugel \emph{et al.} (FINeSSE Collaboration), arXiv:hep-ex/0402007; C. Athanassopoulos \emph{et al.} (LSND Collaboration) Phys. Rev. C \textbf{55}, 2078 (1997); C. Athanassopoulos \emph{et al.} (LSND Collaboration) Phys. Rev. C \textbf{56}, 2806 (1997); V.~Lyubushkin {\it et al.}  [NOMAD Collaboration],  Eur.\ Phys.\ J.\ C {\bf 63}, 355 (2009).

%\bibitem{ncsf61} J.~Grange at NuInt12: \url{https://indico.fnal.gov/getFile.py/access?contribId=83&sessionId=18&resId=0&materialId=slides&confId=5361}.

\bibitem{ncsf63} T.~Golan, Kr.~M.~Graczyk, C.~Juszczak, J.~T.~Sobczyk, Phys. Rev. C \textbf{88}, 024612 (2013).

\bibitem{ncsf64} A.~Bodek, H.~S.~Budd, and M.~E.~Christie, Eur. Phys. J. C \textbf{71}, 1726 (2011).

\bibitem{ncsf65} J.~T.~Sobczyk, Eur. Phys. J. C \textbf{72}, 1850 (2012).

\bibitem{ncsf66} A. M. Ankowski and Omar Benhar, Phys. Rev. D \textbf{88}, 093004 (2013).

\bibitem{ncsf67} M.~V.~Ivanov, A.~N.~Antonov, J.~A.~Caballero, G.~D.~Megias, M.~B.~Barbaro, E.~Moya de Guerra, J.~M.~Udias, Phys. Rev. C \textbf{89}, 014607 (2014).

\bibitem{ncsf79} M.~B.~Barbaro, A.~De~Pace, T.~W.~Donnelly, A.~Molinari, M.~J.~Musolf, Phys. Rev. C \textbf{54}, 1954 (1996).

\bibitem{ncsf70} W.~M.~Alberico, M.~B.~Barbaro, S.~M.~Bilenky, J.~A.~Caballero, C.~Giunti, C.~Maieron, E.~Moya de Guerra and J.~M.~Udias,
  %``Inelastic neutrino and anti-neutrino scattering on nuclei and `strangeness' of the nucleon,''
  Nucl. Phys. A {\bf 623}, 471 (1997).

%\cite{Alberico:1998qw}
\bibitem{Alberico:1998qw}
  W.~M.~Alberico, M.~B.~Barbaro, S.~M.~Bilenky, J.~A.~Caballero, C.~Giunti, C.~Maieron, E.~Moya de Guerra and J.~M.~Udias,  Nucl.\ Phys.\ A {\bf 651} (1999) 277.

\bibitem{ncsf71} M.~C.~Martinez, J.~A.~Caballero, T.~W.~Donnelly and J.~M.~Udias, Phys. Rev. C {\bf 77}, 064604 (2008).

\bibitem{ncsf72} M.~C.~Martinez, J.~A.~Caballero, T.~W.~Donnelly and J.~M.~Udias, Phys. Rev. Lett. {\bf 100}, 052502 (2008).

\bibitem{ncsf73} J.~Liu, R.~D.~McKeown, and M.~J.~Ramsey-Musolf, Phys. Rev. C {\bf 76},
025202 (2007).

\bibitem{ncsf74} R.~Gonzalez-Jimenez, J.~A.~Caballero and T.~W.~Donnelly, Phys. Rep. \textbf{524}, 1 (2013).

\bibitem{ncsf75} E.~L.~Lomon, Phys. Rev. C {\bf 66}, 045501 (2002).

\bibitem{ncsf76} E.~L.~Lomon, Phys. Rev. C {\bf 64}, 035204 (2001).

\bibitem{ncsf77} C. Crawford {\it et al.}, Phys. Rev. C {\bf82}, 045211 (2010).

\bibitem{ncsf78} E. D. Cooper, S. Hama, B. C. Clark, and R. L. Mercer, Phys. Rev. C \textbf{47}, 297 (1993).

\bibitem{RGFa} F.~Capuzzi, C.~Giusti and F.~D.~Pacati, Nucl. Phys. A \textbf{524}, 681 (1991).

\bibitem{RGFb} A.~Meucci, F.~Capuzzi, C.~Giusti and F.~D.~Pacati, Phys. Rev. C {\bf 67}, 054601 (2003).

\bibitem{compcc} A.~Meucci, J.~A.~Caballero, C.~Giusti  and J.~M.~Udias, Phys. Rev. C {\bf 83}, 064614 (2011).

\bibitem{ROP} E. D. Cooper, S. Hama and B. C. Clark , Phys. Rev. C \textbf{80}, 034605 (2009).

\bibitem{ncsf68} D.~Perevalov, ``Neutrino-nucleus neutral current elastic interactions measurement in MiniBooNE'', PhD Thesis, University of Alabama (2009).

\bibitem{ncsf80} \url{http://www-boone.fnal.gov/for_physicists/data_release/ncel}

\end{thebibliography}
\end{document}